\documentclass[fleqn,12pt,twoside]{article}
\usepackage{espcrc1}

\usepackage{graphicx}

\newcommand{\AmS}{{\protect\the\textfont2 A\kern-.1667em\lower.5ex\hbox{M}\kern-.125emS}}
\newcommand{\jj}{$J/\Psi$}

\title{Two-Component Approach to $J/\Psi$ Production in High-Energy
Heavy-Ion Collisions\thanks{This 
work was supported by the U.S. Department of Energy under Grant No. 
DE-FG02-88ER40388.}}

\author{L. Grandchamp\address[SUNYSB]{State University of New York at Stony Brook, 
        Stony Brook, NY 11794-3800}\address{IPN Lyon, IN2P3-CNRS et
        UCBL, 43 Bd. du 11 Novembre 1918, 
        69622 Villeurbanne}\thanks{Email: loic@tonic.physics.sunysb.edu},  
        R. Rapp\addressmark[SUNYSB]\address{NORDITA, Blegdamsvej 17,
          DK-2100 Copenhagen}}
       
\begin{document}
\voffset=-1.2cm
\maketitle
\begin{abstract}
The production of charmonia in ultrarelativistic heavy-ion collisions 
is investigated including two sources. These are a primordial contribution 
coupled with various phases of dissociation, and a statistical coalescence  
of $c$ and $\bar{c}$ quarks at the hadronization phase transition. 
Within a schematic fireball evolution, 
SPS data on \jj\ production can be reasonably well reproduced.  
Remaining discrepancies in the $\Psi'/\Psi$ ratio
are discussed.   Predictions for the \jj\ centrality dependence 
at RHIC energies are confronted with first data from PHENIX.
The pertinent excitation function of the $N_{J/\Psi}/N_{c\bar{c}}$ ratio 
exhibits a characteristic minimum structure signaling the transition from 
the standard \jj\ suppression scenario (SPS) to predominantly 
statistical production (RHIC).
\end{abstract}

\section{Introduction}
In the standard picture of \jj\ production in heavy-ion collisions 
\cite{MS86,Shu80}, charmonium states are (pre-) 
formed in primordial $N$-$N$ collisions and subject to subsequent   
nuclear absorption, Quark-Gluon Plasma (QGP) and hadronic dissociation.
Hence, the magnitude of \jj\ suppression reflects on the plasma effect 
provided that nuclear and hadronic suppression are known and/or small.  
Recently, an alternative mechanism of \jj\ production in relativistic
heavy-ion collisions has been suggested~\cite{GG99,BS00a,GKSG01}. The 
underlying idea is that charmonium states are formed by statistical 
coalescence of $c$ and $\bar{c}$ quarks at the hadronization transition 
according to thermal weights, with the total $c\bar{c}$ abundance being 
determined by primordial parton fusion.  

In the following, we present a combined approach of direct and
statistical \jj\ production~\cite{GR01,GR02}, constructed within a 
thermal framework that is consistent with basic hadro-chemistry and 
expansion dynamics at both SPS and RHIC, and also reproduces 
electromagnetic observables at the SPS. 
In particular, we do not invoke any ``anomalous'' open-charm 
enhancement beyond expectations from $N$-$N$ collision scaling.

\section{Two-component model of \jj\ production}
The first (``direct'') component of \jj\ yield is attributed to 
primordial \jj's arising from early (hard) parton-parton 
collisions. A first suppression 
factor, $\mathcal{S}_{nuc}$, is due to nuclear absorption evaluated in the 
Glauber picture with a phenomenological (constant) cross-section 
$\sigma_{\Psi N} \simeq 6.4$~mb. Subsequent QGP suppression  
is based on inelastic scattering of thermal partons off the $c$ or 
$\bar c$ quark within the bound state, using a quasifree approximation 
with account for in-medium modified charmonium binding energies~\cite{GR01}.
Hadronic dissolution cross sections are estimated within a $SU(4)$-symmetric 
effective theory~\cite{Ha00,LK00}.
The pertinent dissociation rates for both QGP and hadron gas (HG)
are convoluted over a thermal fireball evolution~\cite{RW99,RS00} 
for a heavy-ion collision
at given impact parameter, yielding charmonium suppression factors 
$\mathcal{S}_{QG}^{X}$ and $\mathcal{S}_{H}^X$ ($X$= $J/\Psi$, $\psi'$,
$\chi$), respectively.
This leads to the number of direct
\jj's from primordial production (including feeddown), 
\begin{equation}
  \mathcal{N}_{J/\Psi}^{dir}  = \sigma_{pp}^{J/\Psi}
  ABT_{AB}(b)\mathcal{S}_{nuc}\left[0.6\mathcal{S}_{QG+H}^{J/\Psi} +
    0.032\mathcal{S}_{QG+H}^{\chi}
    +0.08\mathcal{S}_{QG+H}^{\Psi'}\right]  \ .  
\end{equation}

The second (``statistical'') component of \jj\ yield originates from coalescence 
of $c$ and $\bar{c}$ quarks at hadronization. The abundances of charmed 
particles follow from hadronic thermal
weights at the critical temperature $T_c$, with the absolute number
matched to the amount of charm quarks created in primordial 
$N$-$N$ collisions. This necessitates the introduction of an effective
fugacity factor $\gamma_c$ which is determined enforcing exact (local) 
charm conservation. The statistical \jj\ production, including
feeddown,  is then given by 
\begin{equation}
  \mathcal{N}_{J/\Psi}^{th} = \gamma_c^2 V_H
  \left[n_{J/\Psi}\mathcal{S}_H^{J/\Psi}+\sum\mathcal{BR}
    (X\rightarrow J/\Psi)n_X\mathcal{S}_H^X\right]\mathcal{R} \ ,  
\end{equation}
where the volume at hadronization, $V_H$, is taken from our fireball evolution. 
The factor $\mathcal{R}$$\le$1 approximates the effects of incomplete
charm-quark thermalization (calculated in a relaxation time approach), 
inducing a relative reduction of statistical production. 

The total number of \jj's observed per heavy-ion collision follows as  
$\mathcal{N}_{J/\Psi} = \mathcal{N}_{J/\Psi}^{dir} +\mathcal{N}_{J/\Psi}^{th}$.  
The model involves essentially two parameters: the strong coupling constant 
figuring into the QGP dissociation cross section, as well as the 
$c$-quark thermalization time.

\section{SPS}
\begin{figure}[htb]
\begin{minipage}[t]{78mm}
\includegraphics*[width=0.999\textwidth,height=5.78cm]{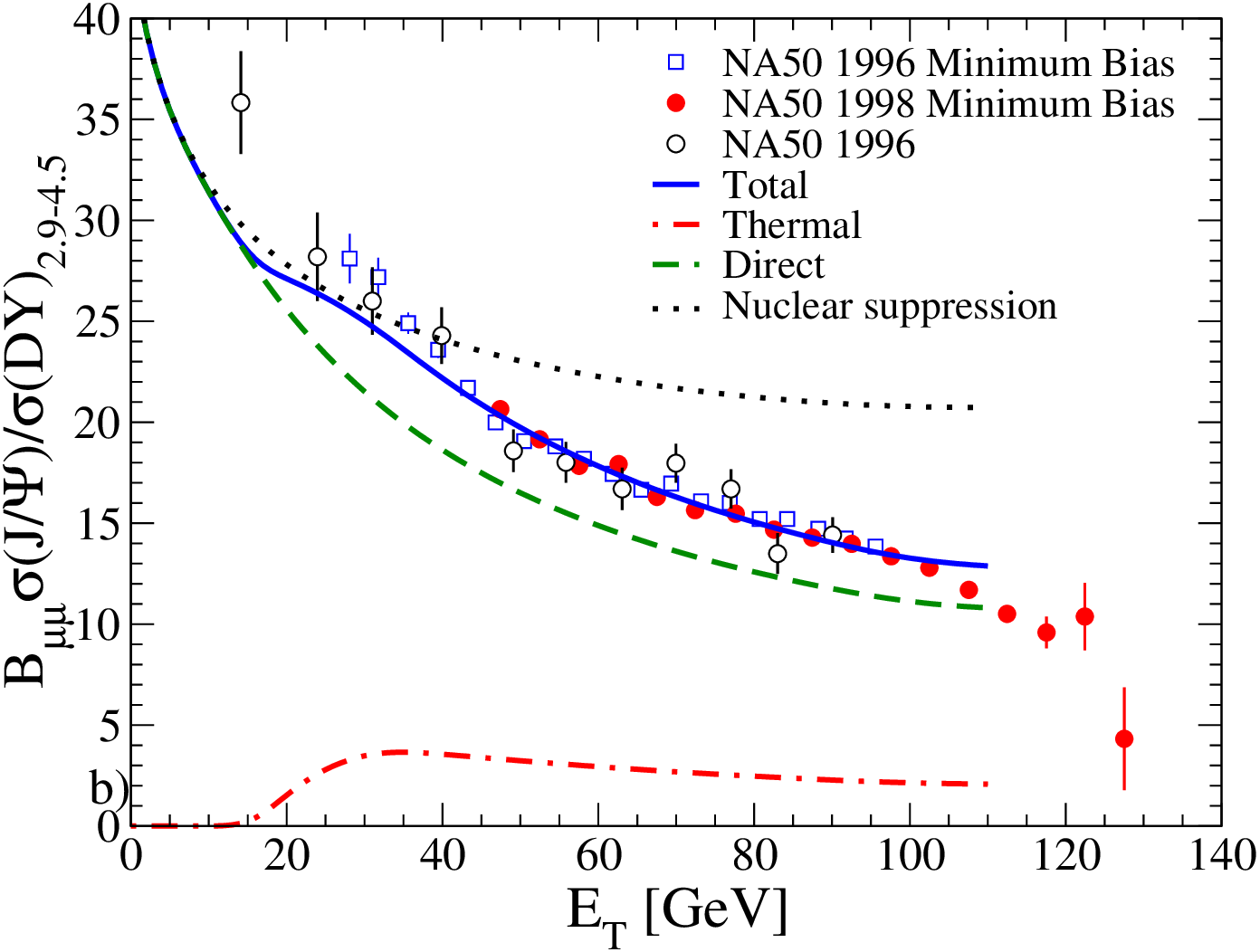}
\vspace*{-1.5cm}
\caption{\jj\ over Drell-Yan ratio at SPS
  for $Pb$(158~AGeV)-$Pb$ collisions.}
\label{fig:SPS}
\end{minipage}
\hspace{\fill}
\begin{minipage}[t]{78mm}
\includegraphics*[width=0.999\textwidth]{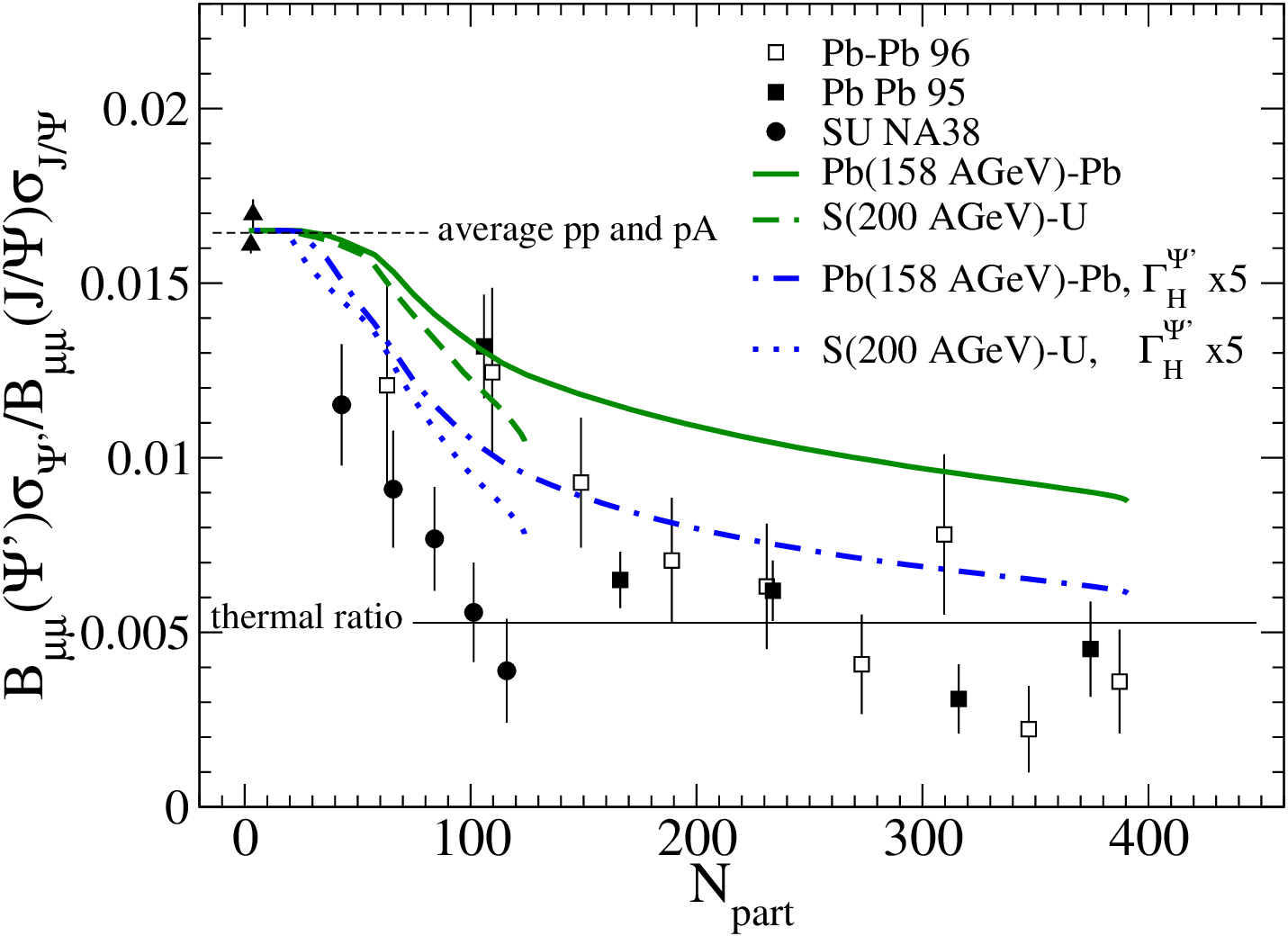}
\vspace*{-1.5cm}
\caption{$\Psi'/\Psi$ ratio for 
  $Pb$(158~AGeV)-$Pb$ and $S$(200~AGeV)-$U$ systems at SPS.}
\label{fig:psip}
\end{minipage}
\vspace*{-0.5cm}
\end{figure}

Our results for the centrality dependence of the \jj\ yield in the 
$Pb$(158~AGeV)-$Pb$ system are compared to NA50 data~\cite{na50-00a} 
in Fig.~\ref{fig:SPS}.  Direct production (dashed line) always prevails
over the thermal component (dot-dashed line). The latter sets in once
a QGP starts forming, which, coupled with the rather strong QGP suppression,
mimics the first drop at around $E_T$$\simeq$40~GeV. The
overall good description of the NA50 data extends 
beyond $E_T>100$~GeV (not shown here) when incorporating transverse energy 
fluctuations~\cite{Cap00} and a careful assessment of the minimum bias 
analysis~\cite{Cap02}, cf.~ref.~\cite{GR02}. In addition, the model also 
reproduces the NA38 results for the $S$-$U$ system reasonably well. 
We note that the relative smallness of statistical production stems
from {\em not} invoking any ``anomalous'' open-charm enhancement.

Within our two-component model, the $\Psi'/\Psi$ ratio can be extracted 
without further assumptions. The results (Fig.~\ref{fig:psip}) 
for both $S$-$U$ (dashed) and $Pb$-$Pb$ (solid) indicate significant 
discrepancy with NA38/NA50 data. We believe the deficiency in our description 
to reside in the hadronic phase, cf.~also ref.~\cite{SSZ97}. Indeed, an artificial
increase of the hadronic $\Psi'$ dissociation cross-section by a factor of 5 
clearly improves the agreement with the data. Further analysis 
is required by addressing, {\it e.g.}, in-medium 
effects on $D$-mesons.

\section{Predictions for RHIC and excitation function}

\begin{figure}[!b]
\vspace*{-0.2cm}
\begin{minipage}[t]{78mm}
\includegraphics*[width=0.999\textwidth]{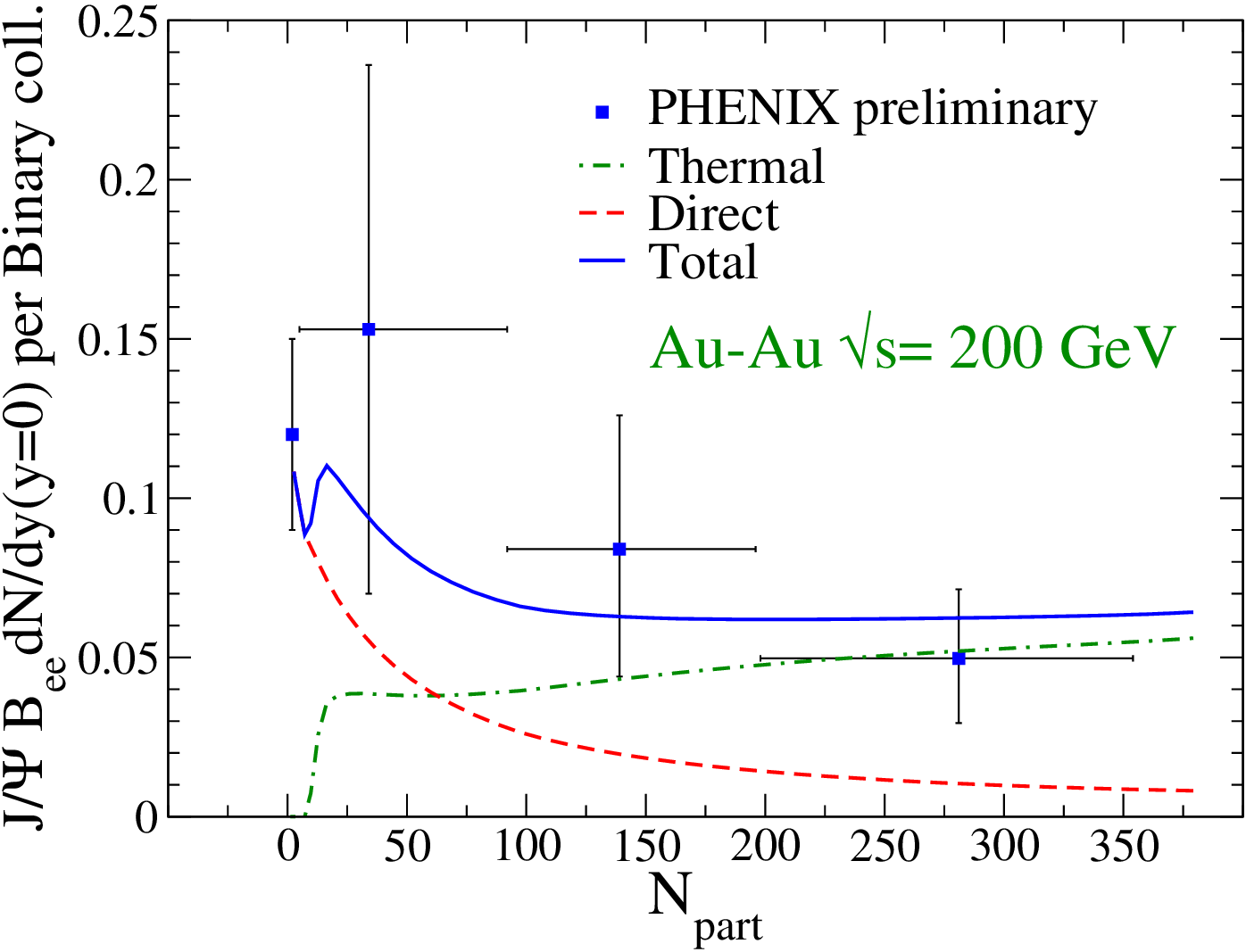}
\vspace*{-1.5cm}
\caption{RHIC centrality dependence of the number of \jj's per
binary collision.}
\label{fig:RHIC}
\end{minipage}
\hspace{\fill}
\begin{minipage}[t]{78mm}
\includegraphics*[width=0.999\textwidth,height=5.94cm]{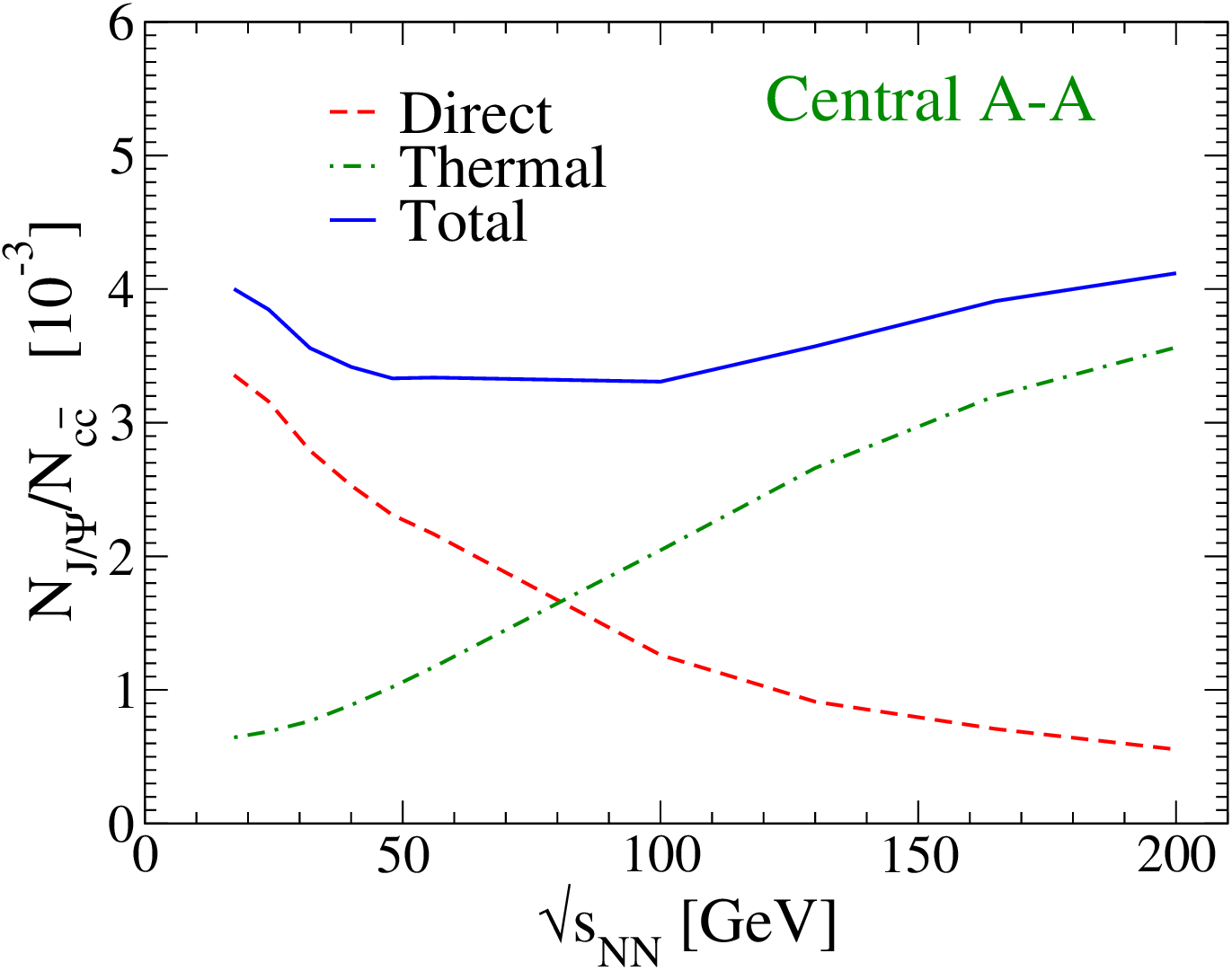}
\vspace*{-1.5cm}
\caption{Excitation function of the ratio $N_{J/\Psi}/N_{c\bar{c}}$
in central heavy-ion collisions.}
\label{fig:excitation}
\end{minipage}
\vspace*{-0.5cm}
\end{figure}

At RHIC, due to much more abundant primordial charm-quark production and
due to a stronger plasma suppression, the statistical \jj\ production
prevails over the direct component. 
This is apparent from Fig.~\ref{fig:RHIC} where the number of \jj's  per
binary collision is plotted as a function of centrality. 
The full curve represents our {\it prediction} of ref.~\cite{GR02}, 
with a minor adjustment of the input parameters from $pp$ collisions
according to the (central values of the) recent PHENIX 
measurements~\cite{FA02} which reported 
$\sigma_{J/\Psi}^{pp}(\sqrt{s}\!\!=\!\!200~\mbox{GeV})
=3.8\pm0.6\mbox{(stat)}\pm1.3\mbox{(sys)}\;\mu$b and
$\sigma_{c,\bar{c}}^{pp}(\sqrt{s}\!\!=\!\!200~\mbox{GeV})\simeq650\;\mu$b. 
Despite the large error bars, it appears that scenarios involving 
standard \jj\ suppression only, as well as
more extreme recombination scenarios, are disfavored by the data. 

We also note that the probabilistic treatment of nuclear absorption,
valid in the low energy regime, is not justified at RHIC energies. 
However, evaluations of nuclear absorption at collider energies using 
a more rigorous quantum-field theoretical approach~\cite{BPSAC98}
find that the suppression effects are quantitatively very close to the 
ones in the probabilistic treatment.

The transition between the regimes of predominantly direct (SPS) to
statistical (RHIC) charmonium production can be further scrutinized 
experimentally at RHIC in terms of an excitation function for the ratio 
$N_{J/\Psi}/N_{c\bar{c}}$, cf.~Fig.~\ref{fig:excitation}.  
This ratio exhibits a nontrivial shallow minimum structure,
a direct consequence of the
interplay between suppressed and thermal production (assuming no anomalies
in open-charm production).

\section{Conclusions}
We have shown that a two-component model for charmonium 
production in heavy-ion collisions, including both ``direct'' \jj's
arising from primordial $N$-$N$ collisions    
subject to nuclear, QGP and hadronic suppression, as well as
``thermal'' \jj's emerging from coalescence of $c$ and $\bar c$
quarks at hadronization, accounts well for \jj\  
centrality dependencies measured at SPS. 
A potential discrepancy has been identified in the $\Psi'/\Psi$ ratio, 
which might hint at shortcomings in the  
evaluation of hadronic $\Psi'$ dissociation. 
Extrapolating our approach to higher energies, the centrality dependence of
\jj\ production at RHIC is in line with preliminary PHENIX data.
We furthermore emphasized 
the importance of an excitation function to determine a potential   
transition from (predominantly)
direct to (mostly) statistical production, as indicated by  
an almost flat (but nontrivial) structure of  
the $N_{J/\Psi}/N_{c\bar{c}}$ ratio as a function collision energy.

\end{document}